\documentstyle[12pt]{article}
\begin{document}
\title{Perfectly secure  cipher system.} \author{Arindam Mitra
\\Lakurdhi, Tikarhat Road, Burdwan. 713102.  India.\\
}
\date{}
\maketitle
\begin{abstract}\bf
We present a perfectly secure cipher system based on the concept
of fake bits which has never been used in either classical or quantum cryptography.

\end{abstract} 
\newpage
\section*{}
Cryptography, the art of secure communication has been
developed since the dawn of human civilization, but it
has been  mathematically treated by Shannon [1].
At present, we have different classical
cryptosystems  whose merits and demerits are discussed below.\\

Vernam cipher [2]: It is proven secure [1] but it can not produce
more absolutely secure bits than the shared secret bits.
Due to this difficulty, it has not become popular, 
however it is still routinely used in diplomatic secure 
communication.\\

Data encryption standard [3] and public key distribution system [4]:
These are widely used cryptosystems because they can produce
more {\em computationally secure} bits than the shared secret bits.
The problem is that its computational security is not proved.
The assumption of computational security has now become 
weak after the discovery of fast quantum algorithms (see ref. 16)\\

To solve the above problems of classical cryptosystem, quantum
cryptography [5-9] has been developed over the last two decades.
Conceptually  quantum cryptography is elegant and many undiscovered
possibilities might store in it. In the last few years work on its
security has been remarkably progressed [10-14], however work is yet not finished.\\
Recently it is revealed [15] that all practical quantum cryptographic
systems are insecure.\\ 

Regarding quantum cryptosystems, the popular  conjectures are:\\
1. Completely quantum channel based cryptosystem is impossible [16] 
(existing quantum cryptosystem requires classical channel to operate).
2. Unconditionally secure quantum bit commitment is impossible [16].
3. By classical means, it is impossible to create more absolutely secure bits
than the shared secret bits.\\

Recently alternative quantum cryptosystem has been developed [17-20] by the
present author; which can operate solely on quantum channel (both entangled
and unentangled type)[17,18] and can provide unconditionally secure
quantum bit commitment [19]. Here we shall see that third conjecture is also
not true.\\  

\noindent
{\bf Operational procedure}: For two party protocol,
the problem of Vernam ciper (popularly called one time pad) [2] is that 
two users have to meet  at regular interval to exchange the key material.
We observe that  key material can be simply transmitted without
compromising security.\\
 
In the presented  cipher system, always in the string of random bits, there
are real and pseudo-bits (fake bits). Real bits contain key material and pseudo-bits
are to mislead eavesdropper. Sender encodes the sequence of real bits
on to the fixed real bit positions and encodes the sequence of pseudo-bits
on to the fixed pseudo-bit positions. It thus forms the entire encoded 
sequence.
which is transmitted. The fixed positions of real and pseudo-bits are
initially secretly shared between sender and receiver. Therefore,
receiver can decode the real bits (the first key) 
from real bit positions. Obviously
he/she ignores the pseudo-bits.\\

For the second encoded sequence, sender uses new sequence 
of real and pseudo-bits
but the position of real and pseudo-bits are same. So again receiver
decodes the second key  from the same real bit positions. In this way
infinite number of keys can be coded and decoded. Notice that
initially shared secret positions of real and pseudo-bits  are
repeatedly used. That's why, in some sense, secrecy is being
amplified. Let us illustrate the procedure.\\
 
\paragraph*{}
\begin{eqnarray}\left(\begin{array}{ccccccccccccccccc}
P & R     & R     & R     & P & R     & P & R     & P &  P & P & R      & P & R     & R     & P &  ....\\
0 & b_{1} & b_{1} & b_{1} & 1 & b_{1} & 1 & b_{1} & 1 &  0 & 0 & b_{1}  & 0 & b_{1} & b_{1} & 1 &  ....\\
1 & b_{2} & b_{2} & b_{2} & 0 & b_{2} & 0 & b_{2} & 0 &  1 & 1 & b_{2}  & 1 & b_{2} & b_{2} & 0 &  ....\\
1 & b_{3} & b_{3} & b_{3} & 1 & b_{3} & 1 & b_{3} & 1 &  0 & 0 & b_{3}  & 0 & b_{3} & b_{3} & 0 &  ....\\
0 & b_{4} & b_{4} & b_{4} & 1 & b_{4} & 0 & b_{4} & 1 &  0 & 1 & b_{4}  & 1 & b_{4} & b_{4} & 0 &  ....\\
. & . & . & . & . & . & . & . & . &  . & . & .  & . & . & . & . &  ....\\ 
. & . & . & . & . & . & . & . & . &  . & . & .  & . & . & . & . &  ....\\ 
. & . & . & . & . & . & . & . & . &  . & . & .  & . & . & . & . &  ....\\ 
1 & b_{n} & b_{n} & b_{n} & 1 & b_{n} & 0 & b_{n} & 0 &  0 & 1 & b_{n}  & 0 & b_{n} & b_{n} & 1 &  ....
\end{array}\right)
\equiv \left(\begin{array}{c}
S_{s} \\ S_{e1} \\ S_{e2} \\ S_{e3}\\ S_{e4} \\. \\. \\.\\ S_{en}\end{array}
 \right)\nonumber\end{eqnarray}

In the above block, the first row represents the sequence
$S_{s}$, which is initially secretly shared. In that sequence,
"R" and "P" denote the position of real and pseudo- bits respectively.
The next rows represent the encoded sequences :
$S_{e1}, S_{e2}, S_{e3}, S_{e4},....,S_{en}$.
In these encoding, $b_{i}$ are
the real bits  for i-th real string of bits.
Other bits are pseudo-bits. 
Obviously the sequences of real bits always form new real keys.
Similarly sequences of pseudo-bits always form new pseudo-keys.
But positions of real and pseudo-bits are unchanged. As 
receiver ignores 
pseudo-bits and pseudo-keys, so the decoded 
strings of real bits (keys) will look like: \\

\paragraph*{}
\begin{eqnarray}\left(\begin{array}{ccccccccc}
 R     & R     & R      & R      & R      & R      & R     & R      &  ....\\
 b_{1} & b_{1} & b_{1}  & b_{1}  & b_{1}  &  b_{1} & b_{1} & b_{1}  &  ....\\
 b_{2} & b_{2} & b_{2}  & b_{2}  & b_{2}  & b_{2}  & b_{2} & b_{2}  &   ....\\
 b_{3} & b_{3} & b_{3}  & b_{3}  & b_{3}  & b_{3}  & b_{3} & b_{3}  &  ....\\
 b_{4} & b_{4} & b_{4}  & b_{4}  & b_{4}  & b_{4}  & b_{4} & b_{4}  &  ....\\
 . &  . & . & .  & . & . & . & . &  ....\\ 
 . &  . & . & .  & . & . & . & . &  ....\\ 
 . &  . & . & .  & . & . & . & . &  ....\\ 
 b_{n} & b_{n} & b_{n} & b_{n}  & b_{n} & b_{n}   & b_{n} & b_{n} &  ....
\end{array}\right)
\equiv \left(\begin{array}{c}
S_{s} \\ K_{1} \\ K_{2} \\ K_{3}\\ K_{4} \\. \\. \\.\\ K_{n}\end{array}
 \right)\nonumber\end{eqnarray}
Here $K_{1}, K_{2}, K_{3}, K_{4}.....,K_{N}$ are 
independent keys.\\ 

Condition for absolute security: Shannon's condition for
absolute security [1] is that  eavesdropper
has to depend on guess for absolutely secure system. In our system, for a
particular encoded sequence of events (bits), if the probability of
real events ($p_{real bits}$) becomes equal to the 
probability of pseudo-events
($p_{pseudo-bits}$) then eavesdropper has to guess. 
Since all the encoded sequences are
independent so eavesdropper has to guess all sequences. 
Therefore, condition for absolute security can be written as:
1.$ p_{pseudo-bits}\geq p_{real bits}$.
2. All encoded sequences should be statistically independent.
 That is, any encoded sequence  should not have pseudo randomness. \\

\noindent 
{\bf Speed of communication:} If we take
$ p_{pseudo-bits}= p_{real bits}$ and share 100 bits,
then  message can be  communicated with 1/4  speed of
digital communication 
(data rate will reduce a factor of 
1/2  due to key production and another factor of 1/2  due to
message encoding) as long as we wish.
 If the key ($K_{i}$) itself
carries meaningful message, then speed of secure communication will 
be just half
of the speed of digital communication. 
Perhaps no cryptosystems offer such speed. \\

The above 
art of key exchange is mainly based on 
the idea of  pseudo-bits, which was first introduced in our
 noised based cryptosystem[21]. But that system will be slow and complicated.
In contrast, this system will be   fast and simple. Note that,
noise has never been a threat to the security of any classical
cryptographic protocol ( rather it can be helpful [21] to achive security). 
This is the main advantage
of classical cryptographic protocol over quantum 
key distribution protocols, where noise indeed a threat to the security.
It should be
mentioned that the  classical cipher system  can not
achieve other quantum cryptographic tasks such
as cheating free Bell's inequality test [18] and
quantum bit commitment encoding [19]. Indeed classical cryptography 
can not be encroach entire area of quantum cryptography.


\begin{thebibliography}{99}
\bibitem{ari27}Mitra. A, Unconditionally secure quantum bit commitment is simply possible.
physics/0007089.
\end{thebibliography}
\end{document}